1:1 alkali-TCNQ salts and the bond order wave (BOW) phase of half-filled linear Hubbard-type models

Z G. Soos[a,*], M. Kumar[a,b], S. Ramasesha[b], and R.A. Pascal, Jr.[a,c]
[a] *Department of Chemistry, Princeton University, USA*
[b] *Solid State and Structural Chemistry Unit, Indian Institute of Science, India*
[c] *Department of Chemistry, Tulane University, USA*

Corresponding author. Tel. 0016092583931; fax: 0016092586746
E-mail address: *soos@princeton.edu* (Z.G. Soos)

**Abstract**
The bond order wave (BOW) phase of half-filled linear Hubbard-type models is narrow and difficult to characterize aside from a few ground state properties. The BOW phase of a frustrated Heisenberg spin chain is wide and tractable. It has broken inversion symmetry $C_i$ in a regular array and finite gap $E_m$ to the lowest triplet state. The spin-BOW is exact in finite systems at a special point. Its elementary excitations are spin-1/2 solitons that connect BOWs with opposite phase. The same patterns of spin densities and bond orders appear in the BOW phase of Hubbard-type models. Infrared (IR) active lattice phonons or molecular vibrations are derivatives of P, the polarization along the stack. Molecular vibrations that are forbidden in regular arrays become IR active when $C_i$ symmetry is broken. 1:1 alkali-TCNQ salts contain half-filled regular TCNQ⁻ stacks at high temperature, down to 100 K in the Rb-TCNQ(II) polymorph whose magnetic susceptibility and polarized IR spectra indicate a BOW phase. More complete modeling will require explicit electronic coupling to phonons and molecular vibrations.

*Keywords:* Extended Hubbard model, Bond order wave phase, TCNQ salts, Frustrated spin-1/2 chains, Infrared active molecular vibrations.

**1. Introduction**

Face-to-face stacks of organic π-radicals suggest a one-dimensional (1D) electronic structure [1,2]. It was realized [1] early on that 1D Hubbard models rationalize the striking magnetic, optical and electronic properties of π-radical solids in terms of the LUMO of π-acceptors A or the HOMO of π-donors D. Segregated stacks of A or D can have different degree of filling by electrons or holes and different unit cells along the stack. Half-filled systems with one electron per site are Mott insulators. Good conductors deviate from half filling in a regular array. Triplet spin excitons signal dimerized stacks. Mixed DA stacks have alternating site energy ±ε and undergo a neutral-ionic transition (NIT) in charge-transfer (CT) salts of special interest. The bandwidth of π-stacks is 4t ~ 1 eV, an order of magnitude smaller than in conjugated polymers with π-bonds along the backbone. 1D quantum cell or Hubbard-type models for π-radical organic solids or conjugated polymers have been extensively developed and applied for decades, sometimes almost quantitatively [3-7]. Both electron-phonon coupling and electron correlation are important.

The versatile π-acceptor A = TCNQ (tetracyanoquinodimethane) was the focus of early studies. TCNQ forms segregated stacks with many closed-shell cations and CT salts with strong π-donors. Vegter and Kommandeur [8] reported the magnetism and dimerization transition at $T_d$ of 1:1 salts M-TCNQ, where M = Na, K, Rb or Cs. These salts have half-filled regular A⁻ stacks for T > $T_d$. They are prime candidates for modeling, especially the polymorph Rb-TCNQ(II) that has regular stacks [9] at 100 K instead of [8] $T_d$ ~ 220 K.

Models have to be understood before they can be applied, however, and correlated 1D models are very challenging. Several theoretical advances make it timely to study 1:1 alkali-TCNQ salts again. The first is the phase diagram of half-filled regular arrays. Nakamura [10] pointed out that the extended Hubbard model (EHM) has a narrow bond order wave (BOW) phase around $V \sim U/2$, where $U > 0$ is the on-site repulsion and V is the nearest-neighbor Coulomb interaction. The ion-radical stack $(A^-)_n$ is stable up to $V = U/2$ when $t = 0$; larger V yields a charge density wave (CDW) with alternating $A^{-2}$ and A along the stack. The second is the modern theory of polarization [11] and its Berry-phase formulation [12,13]. It is essential for rigorous vibrational analysis of models, as illustrated in mixed-stack NIT systems [14,15]. Regular $(A^-)_n$ arrays have electronic coupling to lattice phonons such as the Peierls mode and to totally symmetric (ts) molecular vibrations that modulate $\epsilon$. The third is the role of topological solitons or domain walls introduced by Su, Schrieffer and Heeger (SSH) for polyacetylene [6] and generalized to other 1D models [7,16]. Last but not least, advances such as the density matrix renormalization group (DMRG) [17] have vastly improved direct analysis.

The BOW phase of the EHM is poorly characterized aside from some key aspects: [10,18-21] a doubly degenerate ground state (gs), broken inversion symmetry $C_i$ in a regular array and a finite energy gap $E_m$ to the lowest triplet state. These properties are retained in a spin chain with a BOW phase that can be studied in detail, as summarized in Section 2. The Berry-phase formulation of polarization is applied in Section 3 to systems with broken electronic $C_i$ symmetry, charge fluctuations and linear coupling to the Peierls phonon $\delta$ and a ts molecular vibration $\Delta$. Such a correlated 1D model is minimally required for half-filled segregated stacks in M-TCNQ salts.

## 2. BOW phase of spin or Hubbard models
We consider a regular array of $s = 1/2$ sites with Heisenberg antiferromagnetic (HAF) exchange $J_1 = J(1 - x)$ between first neighbors and $J_2 = xJ$ between second neighbors. The Hamiltonian with frustrated exchange is

$$H(x) = J\sum_n \left((1-x)\vec{s}_n \cdot \vec{s}_{n+1} + x\vec{s}_n \cdot \vec{s}_{n+2}\right) \tag{1}$$

Majumdar and Ghosh [22] identified the special point $J_2 = J_1/2$, or $x = 1/3$. The exact gs of $H(1/3)$ for an even number of sites N and periodic boundary conditions (PBC) is one of the Kekulé diagrams of organic chemistry,

$$\begin{aligned}|K1\rangle &= (1,2)(3,4).....(N-1,N) \\ |K2\rangle &= (2,3)(4,5).....(N,1)\end{aligned} \tag{2}$$

The notation (n,n+1) indicates singlet paired spins, $(\alpha_n\beta_{n+1} - \beta_n\alpha_{n+1})/\sqrt{2}$, at adjacent sites. $|K1\rangle$ is the nondegenerate gs [23] for open boundary conditions (OBC), and the energy is $-NJ/4$ in either case.

The BOW phase of H(x) starts at the fluid-dimer transition at $J_2/J_1 = 0.2411$ ($x_1 = 0.1943$) found by Okamoto and Nomura [24]. At $x = 1/3$, finite PBC systems illustrate BOWs, broken $C_i$ symmetry, and finite $E_m$. The BOW amplitude is $B = 3/8$, since $-\langle \vec{s}_n \cdot \vec{s}_{n'}\rangle = 3/4$ when sites n, n' are paired and zero otherwise. The excitation energies of H(1/3) can be found [23] directly up to $N \sim 30$. Finite-size effects are small for the triplet $E_m$ or first excited singlet $E_3$; they are larger for the lowest quintet $Q_1$ or for higher triplets and singlets.

$|K1\rangle$ or $|K2\rangle$ is also a convenient representation of partial double and single $\pi$-bonds in linear polyenes or in polyacetylene. In contrast to such *dimerized* systems, however, $|K1\rangle$ is the gs of a *regular* array of spins. The SSH model [25,6] describes topological solitons or domain walls between $|K1\rangle$ and $|K2\rangle$ regions in polyacetylene. We model spin solitons of H(1/3) in a regular array with OBC and odd $N = 4n \pm 1$. The gs is a doublet $S = S_z = 1/2$. The spin densities $\rho_p = 2\langle S_{pz}\rangle$ for $N = 25$ and 23 are shown in Fig. 1 (top panel) with $p = 0$ at the center and $p = \pm 2n$ or $\pm 2(n-1)$ at the ends. They are typical $\rho_p$ in dimerized correlated systems [26], with $\rho_p < 0$ at sites where SSH solitons have $\rho_p = 0$.

The bond orders $-\langle \vec{s}_p \cdot \vec{s}_{p+1} \rangle$ show almost perfect pairing at either end; as expected, they decrease and reverse at the middle [23]. The energy $2E_W$ for two solitons is

$$2E_W(N) = 2E_0(N) - E_0(N+1) - E_0(N-1) \qquad (3)$$

with odd N. At x = 1/3 we obtain good size convergence, $2E_W/J$ = 0.1701, 0.1684 and 0.1669 for N = 23, 25 and 27. As expected, we have $2E_W(N) \sim E_m(N+1) \sim E_3(N+1)$. The lowest triplet or singlet excited state for even N and PBC is a pair of solitons [23]. Spin-1/2 solitons are the elementary excitations not only of H(1/3), but over the entire BOW phase from $x_1$ = 0.1943 to $x_2$ = 0.67. The soliton width is $2\xi \sim 15$ at x = 1/3.

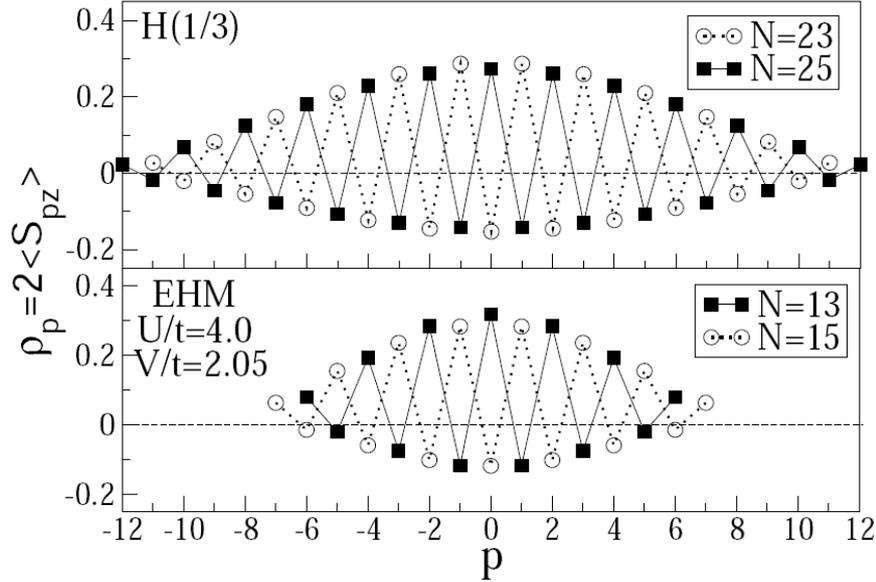

Fig. 1. BOW phase spin densities $\rho_p = 2\langle S_{pz} \rangle$ of linear N-site radicals with p = 0 at the center and p = ±2n or ±2(n−1) at the ends. Top: Spin chain H(1/3) for N = 23 and 25. Bottom: EHM with U = 4t, V = 2.05t for N = 13 and 15. The N = 4n+1, 4n−1 series have large $\rho_0 > 0$, small $\rho_0 < 0$, respectively.

Hubbard models also have charge degrees of freedom. The EHM with t = 1.0, U = 4.0 and V = 2.05 is in the BOW phase [1,21], although the lowest singlets $\psi_{\pm 1}$ with $C_i = \pm 1$ are not quite degenerate at N = 12 or 16; they differ by less than 0.005t per site. Neglecting these finite-size effects, we construct,

$$|\psi_\pm\rangle = (|\psi_1\rangle \pm |\psi_{-1}\rangle)/\sqrt{2} \qquad (4)$$

$\psi_\pm$ are analogous to the broken-symmetry Kekulé diagrams of the spin chain at x = 1/3, whose even and odd linear combinations transform as $C_i = \pm 1$. Bond orders $p_r$ are now gs expectation values

$$2p_r \equiv \left\langle \sum_\sigma \left(a_{r\sigma}^+ a_{r+1\sigma} + hc\right) \right\rangle = \frac{\partial E_0}{\partial t_{r,r+1}} \qquad (5)$$

$a_{r\sigma}^+$ ($a_{r\sigma}$) creates (annihilated) an electron with spin σ at site r and hc is the Hermitian conjugate. The $\psi_\pm$ expectation values for the EHM with PBC, N = 16 and parameters above return $p_+$ = 0.752, $p_-$ = 0.398. The broken-symmetry states have $p_+$ for bonds 2r−1, 2r or 2r, 2r+1, respectively. The actual gs is $\psi_{-1}$ and has uniform p = 0.5742 that is within 1% of $(p_+ + p_-)/2$. Charge fluctuations are large [27] since p is close to the band limit of $2/\pi$ = 0.637 for free electrons. Results for N = 12 point to modest (5-10%) and opposite finite-size effects for $p_\pm$. Finite-size effects for p are only 1%. The BOW phase of Hubbard-type models is close to the metallic point that marks a continuous BOW-CDW boundary.

The BOW phase of the EHM or of other spin-independent potentials [21] describes a regular array with one electron per site. The electronic structure of a rigid TCNQ⁻ stack is such a system for microscopic parameters that return a BOW phase. The spin densities $\rho_p$ in Fig. 1(bottom panel) are for N = 13 and 15 with OBC for the EHM with the above parameters. Aside from the restriction to smaller N, the $\rho_p$ pattern is similar and has negative $\rho_p$ that indicate AF correlations. Also as in the spin chain, the terminal bond orders are close $p_+$, and they decrease and reverse in the middle. Broken electronic symmetry in the BOW phase of spin or Hubbard models leads to properties that resemble broken $C_i$ symmetry due to dimerization.

### 3. Infrared active molecular vibrations

The polarization P per unit length and charge of an extended 1D insulator can formally be written as [12]

$$P = (2\pi)^{-1} \operatorname{Im} \ln \langle \psi_0 | \exp(2\pi i M/N) | \psi_0 \rangle \tag{6}$$

$\psi_0$ is the exact gs of an N-site supercell with PBC and M is the conventional dipole operator. The restriction to insulators is [13]

$$|Z| \equiv |\langle \psi_0 | \exp(2\pi i M/N) | \psi_0 \rangle| \neq 0 \tag{7}$$

$Z(\delta_e,\varepsilon)$ is readily obtained for N-site correlated 1D models whose equilibrium structure has alternating $t = -(1 \pm \delta_e)$ and energy $\pm\varepsilon$ along the chain [14,15]. The dipole operator is $M = \Sigma_r r(n_r - z_r)$ where $n_r$ is the number operator for electrons and $z_r$ is the charge when $n_r = 0$. The condition $|Z| \neq 0$ excludes a 1D band with $\delta_e = \varepsilon = 0$. The Berry-phase expression for P makes possible a rigorous treatment of polarization in extended systems by making the dipole M compatible with PBC.

IR activity polarized along the chain is given by derivates [15] such as $(\partial P/\partial Q)_e$ where Q is a k = 0 lattice phonon or ts molecular vibrations. Molecular vibrations that modulate site energies as $\pm\Delta$ generate a dipole $\mu(\delta_e)$ along the stack,

$$\mu(\delta_e) = \left(\frac{\partial P(\delta_e,\Delta)}{\partial \Delta}\right)_\varepsilon = \frac{1}{2\pi}\left(\frac{\partial}{\partial \Delta}\frac{\langle \sin M' \rangle}{\langle \cos M' \rangle}\right)_\varepsilon \tag{8}$$

where M' = $2\pi M/N$ and the expectation value is for the gs at $\delta_e$, $\varepsilon$. The last part of Eq. (8) holds when the gs is real, here a linear combination of many-electron VB diagrams with real coefficients.

Since M is odd under inversion, a polar system with finite $P(\delta_e,\varepsilon)$ requires both dimerization and site energies. $C_i$ symmetry at sites insures that $Z(0,\varepsilon)$ is real, and hence $P(0,\varepsilon) = 0$. Mixed DA stacks have finite $\varepsilon$ and show strong IR activity [28] of molecular modes on dimerization at the NIT. The regular stack at T > $T_d$ has a soft Peierls mode $\omega_P$ < 100 cm$^{-1}$ that appears as a combination band [29] in mid-IR spectra and is understood [15] as $(\partial^2 P/\partial\Delta\partial\delta_P)_e$ where $\Delta$ ($\delta_P$) is a molecular (Peierls) mode. In addition, the dispersion of the optical phonon branch is anomalous [30] when $T_d$ is approached from above. $P(\delta_e,\varepsilon)$ works well for CT salts.

Segregated stacks with $\varepsilon = 0$ have $P(\delta_e,0) = 0$ due to inversion symmetry at the center of bonds or electron-hole (eh) symmetry in models. Finite $\mu(\delta_e)$ requires dimerized stacks such as K, Na or Rb-TCNQ(I) at T < $T_d$, and polarized IR spectra yield detailed information [31,32] about coupling to molecular vibrations as well as evidence for dimerization. The regular 100 K structure [9] of Rb-TCNQ(II) and intense IR spectra [33,34] contradict this well-established paradigm. A BOW phase resolves [9] the problem by breaking electronic $C_i$ symmetry without dimerization.

We need $\mu(\delta_e)$ in Eq. (8) in a BOW phase with $\delta_e = 0$. We again take an EHM with U = 4.0 and compute $P(\delta_e,\Delta)$ for N = 12 and 16. The derivate is readily found since $P(\delta_e,\Delta)$ is strictly linear for $\Delta$ < 0.01. The induced dipole $\mu(\delta_e)$ vs $\delta_e$ is shown in Fig. 2 for N = 12 and 16 for different values of V. The fluid-BOW boundary [1,21] of the EHM with U = 4 is V ~ 1.86, just where $\mu(\delta_e)$ develops a weak maximum. Finite-size effects are negligible for V < 1.8 and are still modest at V = 2.05, close to the metallic point at the BOW-CDW boundary. Likewise, finite-size effects for N ~ 16 are small [15] in mixed stacks except close to the NIT. The reason for $\mu(0) = 0$ in Fig. 2 is clearly that the nondegenerate gs for finite N = 4n transforms as $C_i = -1$. The broken-symmetry states $\psi_\pm$ in Eq. (4) require a degenerate gs as found explicitly in the spin chain H(1/3). Small $\pm\delta_e$ at V = 2.05 is sufficient to generate a large dipole. Our interpretation of Fig. 2 is that $\mu(0)$ vanishes in the fluid phase, becomes finite at the fluid-BOW boundary, increases in the BOW phase and then vanishes abruptly at the BOW-CDW boundary. The CDW phase at large V has broken eh rather than $C_i$ symmetry, with occupation number n > 1 on one sublattice and n < 1 on the other. We understand why $\mu(0)$ is finite in the BOW phase, but not yet how it depends on model parameters.

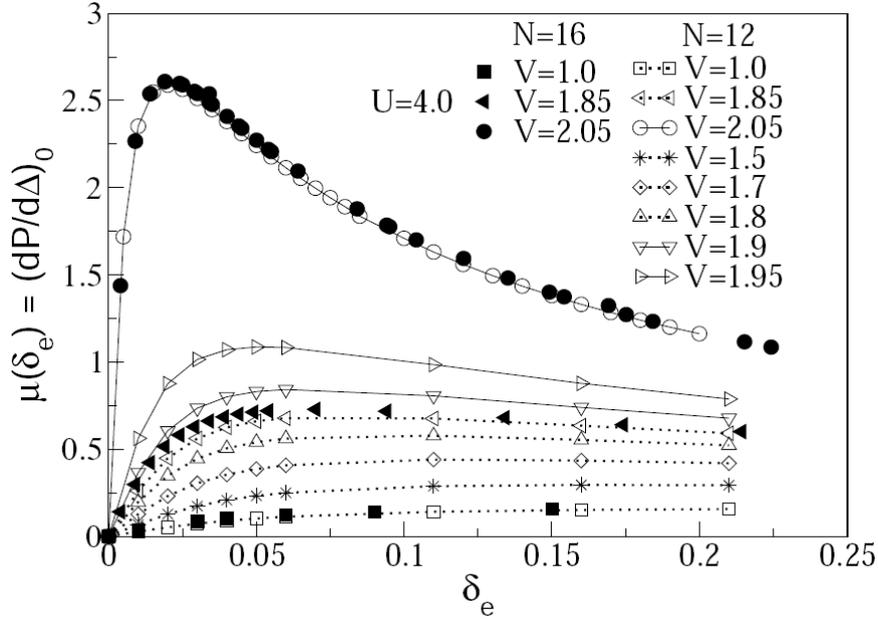

Fig. 2. Dipole $\mu(\delta_e)$ in Eq. (8) along a chain with $t = t(1 \pm \delta_e)$ induced by a totally symmetric molecular vibration in an EHM with N = 16 or 12 and U = 4t for a series of V/t values; $\mu(\delta_e)$ develops a maximum for V > 1.86t (solid lines), the entrance to the BOW phase, that increases up to the metallic point at V = 2.16t.

The intensity $I_{IR}(T)$ of IR active molecular vibrations goes as $\mu(\delta_e)^2$. Dimerized TCNQ$^-$ stacks with $\delta_e \sim 0.2$ are assumed but $\alpha = dt/du$ is needed for an accurate value in addition to the crystal structure. The $\mu(\delta_e)$ dependence in Fig 2 is fairly weak, and certainly much weaker than $1/\delta_e$ in a half-filled band [15] of free electrons with $\varepsilon = 0$. Finite $\mu(0)$ in a BOW phase then rationalizes comparable [34] $I_{IR}(T)$ at low T in the K and Rb(II) salts. Indeed, $I_{IR}(T)/I_{IR}(25K)$ data were used to infer [33] a dimerized Rb-TCNQ(II) structure below 220 K. The 100 K stacks are regular, however, within the limits of a good structural determination [9]. A BOW phase with finite $E_m$ accounts for IR activity as well as negligible spin susceptibility, $\chi_M(T)$, below 150 K.

The cooperative decrease [34,33] of $I_{IR}(T)/I_{IR}(77K)$ up to T = 400 K is another striking feature of Rb-TCNQ(II). The intensity decrease is roughly correlated to increasing $\chi_M(T)$ that we associate with spin-1/2 solitons. Solitons introduce local inversion centers about which bond orders reverse [6]. It follows that IR activity is reduced over a soliton width of $2\xi$ that we cannot evaluate at present, and the structure of an infinite chain with many solitons is not known at finite T even in the band limit. The number density of solitons can be related to the spin density $\rho_s = \chi_M(T)/\chi_{Cu}(T)$ on dividing by the Curie susceptibility. Solitons begin to overlap at $2\xi\rho_s \sim 1$. Very qualitatively, the IR intensity and the relation to $\chi_M$ suggest the following T dependence for induced dipoles,

$$\mu(0,T) = \mu(0)/(1+2\xi\rho_s) \tag{9}$$

Here $\mu(0)$ is the BOW-phase value at $\delta_e = 0$ and increasing $\rho_s(T)$ gives the T dependence that initially goes as $1 - 2\xi\rho_s$. The relative IR intensity of ts molecular modes decreases as $1/(1 + 2\xi\rho_s)^2$. $\chi_M(T)$ data [9] for Rb-TCNQ(II) and $2\xi = 20$ lead to $2\xi\rho_s \sim 1$ at 260K and fit well the relative IR intensity of the 722 cm$^{-1}$ mode in Fig. 2 of [33] between 100 and 300 K, but not the ~5% increase between 100 and 25 K where $\rho_s \sim 0$.

## 4. Discussion and Conclusions

Theoretical studies [10,18-20] of the BOW phase of the EHM did not consider possible physical realizations. Although poorly characterized at present, the BOW phase of half-filled Hubbard-type models addresses [9] the most puzzling Rb-TCNQ(II) data. It combines regular stacks with $C_i$ symmetry at 100 K with negligible $\chi_M(T)$ below 150 K that indicates finite $E_m$ and IR active molecular modes that indicate broken electronic $C_i$ symmetry. The T dependence of IR intensities and $\chi_M(T)$ require excited states that are not yet known. In view of other 1D systems, realistic treatment will have to include linear coupling $\alpha = dt/du$ to lattice phonons and $\Delta$ to ts molecular vibrations.

The frustrated spin-1/2 chain H(x) has a wide BOW range and is particularly simple at x = 1/3. For even N, the exact gs of H(1/3) with PBC is either Kekulé diagram in Eq. (2), and it is $|K1\rangle$ with OBC. The spin chain is far more tractable. Its elementary excitations are spin solitons in a regular array [23] with broken $C_i$ symmetry that mimic many aspects of SSH solitons between regions with opposite dimerization. The spin chain provides insight about excitations and magnetism of BOW systems, just as HAFs approximate magnetism of Hubbard models. Finite $E_m$ ensures an exponentially small $\chi_M(T)$ in the spin-BOW phase at low T, followed by an almost linear increase [23] to a broad maximum. A broad maximum is characteristic of HAFs or Hubbard models while finite $E_m$ signifies a BOW phase.

Rb-TCNQ(II) is a realization of a BOW system [9]. K or Na-TCNQ have smaller but similarly increasing [8] $\chi_M(T)$ at $T > T_d$ that suggest a BOW phase with larger t. All alkali-TCNQ salts are close [27] to the CDW boundary where the 3D electrostatic (Madelung) energy of the crystal is equal to the disproportionation energy $2A^- \rightarrow A + A^{2-}$. Quantum chemical calculation [35] gives larger t for K or Na-TCNQ with almost eclipsed (ring-over-ring) stacks [36] than for Rb-TCNQ(II) with slipped (ring-over-bond) stacks [36]. Larger t also suggests stronger coupling $\alpha = dt/du$ to lattice phonons that may induce dimerization. The K or Na-TCNQ transitions at $T_d$ = 395 or 348 K are not purely 1D, however, since the cation lattice also dimerizes [37]. A BOW phase may allow improved treatment of $T_d$. The broadly similar optical, magnetic and electrical properties of 1:1 alkali-TCNQ salts virtually demand joint modeling.

The BOW phase of Hubbard-type models underscores the importance of understanding 1D models. The best available models were applied imaginatively and speculatively in the exciting early days of TCNQ salts, organic conductors and superconductors. More quantitative modeling is now possible in far more mature fields. The simplicity of 1:1 alkali-TCNQ salts was deceptively attractive. They are not HAFs or spin-Peierls systems or Hubbard models, but are probably BOW systems with intersite interactions, coupling to lattice and molecular vibrations, and 3D Coulomb interactions. There are many open questions about either 1:1 alkali-TCNQ salts or BOW phases.

**Acknowledgments**
ZGS thanks A. Girlando for access to unpublished IR spectra and A. Painelli for discussions of polarization and BOW phases. We thank the National Science Foundation for partial support through the Princeton MRSEC (DMR-0819860).